\documentclass[aps,twocolumn,groupedaddress,showpacs]{revtex4}

\usepackage{graphicx}
\usepackage{epsfig}

\def\begeq{\begin{equation}}
\def\endeq{\end{equation}}

\def\begeqar{\begin{eqnarray}}
\def\endeqar{\end{eqnarray}}

\def\t{\theta}

\def\g{\gamma}

\begin{document}
\bibliographystyle{apsrev}


\title{How Irrelevant Operators affect the Determination of Fractional Charge} 


\author{A.~Koutouza$^1$, H.~Saleur$^{1,2}$, and B.~Trauzettel$^2$}
\affiliation{${}^1$Department of Physics,
University of Southern California, Los Angeles, CA 90089-0484\\
${}^2$Physikalisches Institut, Albert-Ludwigs-Universit\"at,
 D-79104 Freiburg, Germany}


\date{\today}

\begin{abstract}
We show that the inclusion of irrelevant terms in the hamiltonian describing
tunneling between edge states in the fractional quantum Hall effect (FQHE) can 
lead to a variety of non perturbative behaviors in intermediate energy
regimes, and, in particular,  affect crucially the determination of  charge through shot noise measurements. We show, for instance, that certain combinations of relevant and irrelevant terms can lead to  an effective  measured charge $\nu e$ in the
strong backscattering limit and an effective measured charge $e$ in the weak
backscattering limit, in sharp contrast with standard perturbative expectations. 
This provides a possible scenario to explain the  experimental
observations by Heiblum {\em et al.}, which are so far not understood.
\end{abstract}

\maketitle

   
Recent experiments \cite{Comforti} have raised the possibility that Laughlin
quasiparticles (LQPs) \cite{Laughlin} may be able to tunnel through very high
barriers in a FQHE tunneling experiment if the beam of LQPs is diluted. This is unexpected. Conventional wisdom says instead that in the limit of strong pinching,
the quantum Hall fluid is split into two pieces. The vicinity of this limit, if
it can be described perturbatively, can only involve tunneling of
electrons from one half to the other, and noise experiments should
give an effective carrier charge equal to $e$. The experimental observations 
\cite{Comforti} - an effective tunneling charge as low as $0.45 e$
for a measured transparency of only $0.1$ - seems to be
counterintuitive. Furthermore, unpublished experiments by Chung {\em et al.}~\cite{ComfortiI} revealed an effective tunneling charge $e$ for a full beam of
LQPs incident on an almost open quantum point contact. Both of these
surprising observations are so far not understood. In a recent work, Kane and Fisher
\cite{KaneFisher02} investigated by a mix of perturbative and non
perturbative techniques, the tunneling of diluted beams of LQPs in a set up 
similar to the one in Ref.~\cite{Comforti}, and found no evidence for the
tunneling of LQPs through high barriers. In this letter, we consider, using
techniques of integrable field theory, tunneling hamiltonians in a strongly non
perturbative region, where irrelevant terms play a major role. We show that,
as a result  of strong interactions, it is in fact very possible to 
measure an effective tunneling charge $\nu e$ in a region of small
transparency and $e$ in a region of high transparency, a result at odds
with perturbative intuition. We use the electron charge as unit and set $h
\equiv 1$ from now on.

Let us first recall the standard theoretical set-up for this problem.
After bosonizing the chiral edges action, and performing some folding
transformations, the  hamiltonian 
with only the most relevant term included is the boundary sine-Gordon model
\cite{KFold}:  
\begin{equation}
    H={1\over
    2}\int_{-\infty}^{0}\left[(\partial_{x}\phi)^{2}+
    \Pi^{2}\right]+v\cos\sqrt{2\pi\nu}\phi(0).\label{ham}
    \end{equation}
This model is integrable 
\cite{GhoZam}, and DC current and noise can be calculated
    \cite{FLSI,FLSII}. In the weak backscattering limit, obtained at small $v$
    or large energies, the backscattered current is small, and the noise is
    determined by incoherent tunneling of LQPs of charge $\nu$. On the other
    hand, in the strong backscattering limit, the transmitted current is small, and the noise is
determined by incoherent tunneling of electrons. We exclusively
    consider the case of zero temperature, $T=0$, in this note. The
    exact solution of the hamiltonian relies crucially on an integrable
    quasiparticle description. At zero bias, $V=0$, the ground state of the
    theory is just the vacuum with neither kinks nor breathers; the
    quasiparticles are in fact defined as excitations above this vacuum. For $V>0$, kinks of charge unity start filling the vacuum. Unfolding the theory (\ref{ham}) gives rise to right moving particles only, and we parameterize their energy and momentum by $\epsilon=p=e^\theta$, $\theta$ the rapidity. The new ground state is
made of kinks occupying the range $\theta\in (-\infty, A]$;
in other words, the filling fractions are 
$f_+(\t,V)=1$ for $\t<A$ and $f_+(\t,V)=0$
for $\t>A$.
The surface of the sea is approximately
$A \approx \ln (V)$,  but
computing $A$ exactly requires some technology due to the kink interaction.
There are no antikinks nor breathers in the sea at $T=0$, so their densities
do not appear in this analysis. For ease of
notation, we define $\rho(\t)\equiv n_+(\t,V)f_+(\t,V)|_{T=0}$.
At $T=0$ the
quantization of allowed momenta reads:
\begeq
2\pi \rho(\t)=e^\theta
+2\pi\int_{-\infty}^A
\Phi(\theta-\theta')\rho(\theta')d\theta',
\label{dens}
\endeq
where
$\rho=0 \hbox{ in }[A,\infty)$, and $\Phi={1\over 2i\pi} {d\over d\theta}
S_{++}(\t)$ follows from the kink-kink scattering matrix in the bulk. A
straightforward exercise in Fourier transforms and Wiener Hopf integral
equations \cite{FLSbig} leads to $ e^A={V\over 2}{G_+(0)\over G_+(i)}$, 
and thus,
\begeq
\tilde{\rho}(\omega)={1\over 2i\pi} {G_-(\omega)G_+(i)\over \omega-i} e^{(i\omega+1)A}.
\endeq
where the kernels have been defined as ($\g={1\over\nu}-1$)
\begeq
G_+=\sqrt{2\pi(\g+1)}
{\Gamma(-i(1+\g)\omega/2\g)\over
\Gamma(-i\omega/2\g)\Gamma(1/2-i\omega/2)} e^{-i\omega\Delta},
\label{eqv}
\endeq
and $G_-(\omega)=G_+(-\omega)$, $\Delta\equiv{1\over 2}\ln 
\g-{1+\g\over 2\g}\ln(1+\g)$.

It is important to stress that the quantization of the bulk theory (we neglect
finite length effects here) is {\sl independent of 
the boundary interaction}. The mere choice of the 
quasiparticle basis however does depend
crucially on this interaction; in fact, it is chosen precisely in
such a way that the quasiparticles scatter one by one, without
particle production, on the boundary. This is described by a
scattering matrix  element $Q$, in terms of which the current reads \cite{FLSbig}
\begeq
I(V,T_B)=\int_{-\infty}^A \rho(\theta)
\left|Q\left(\t-\t_B\right)\right|^2 d\theta 
\endeq
with $|Q|^{2}= 1/(1+\exp[2(1-1/\nu)(\t-\t_{B})])$. Here, $\theta_B$ is a
measure of the strength of the tunneling term in (\ref{ham}),
$e^{\theta_B}\propto v^{1\over 1-\nu}$. Similarly, the noise can be written as
\begeq
S(V,T_B)=\int_{-\infty}^A \rho(\theta)
\left(\left|Q\left(\t-\t_B\right)\right|^2
-\left|Q\left(\t-\t_B\right)\right|^4\right)d\theta \; .
\endeq
In terms of renormalization group theory (RG), the boundary interaction induces a
flow from Neumann (N) to Dirichlet (D) boundary conditions.  It is widely
believed that these are the only possible conformal boundary conditions 
for the compact boson. The whole operator content at the strong backscattering
(D) fixed point can be determined using
general considerations of conformal invariance \cite{Affleck}: the allowed
operators  have dimension $h_{D}=n^{2}/\nu$ modulo integers, where $n$
is an integer.  We stress that {\sl there 
is no operator of dimension $\nu$}  at this fixed point. Instead, the most 
relevant operator has  $h=1/\nu$. This operator is
$\cos\sqrt{2\pi/\nu}\tilde{\phi}$ ($\tilde{\phi}$ being the dual of
the field $\phi$) and corresponds to transfer of charge unity.

The allowed operators at the weak backscattering (N) fixed point have
dimensions $h_{N}=\nu m^{2}$ modulo integers, where $m$ is an integer. The tunneling of LQPs at this fixed point
corresponds to $m=1$, and to the most relevant operator in its vicinity. Note
that since $\nu$ is the inverse of an (odd) integer, the  operator of
dimension $h=1/\nu$ is allowed near this fixed point. This operator conforms
to tunneling of a bunch of $1/\nu$ LQPs -- in a sense an electron -- in the weak backscattering limit.  

The two types of tunneling charges can be identified theoretically
and experimentally using the noise. Near the high energy fixed point, the
noise as well as  the backscattered current are due to rare tunnelings of
LQPs, so $S\approx \nu (I_0 - I)$, while near the low energy fixed point, the
noise as well as  the current are due to tunneling of electrons, so
$S\approx I$. These results may appear to contradict the fact that in the
integrable approach, the quasiparticles always have charge unity. What happens
of course is that in the weak backscattering limit, these quasiparticles
have non negligible probabilities of tunneling, so the term $|Q|^4$ is of the
same order than $|Q|^2$, resulting into a slope $(-\nu)$. 

Interestingly, hamiltonians with irrelevant terms added to the basic tunneling
term in (\ref{ham}) are also solvable. Some technical points of such a model
were already discussed in the case $\nu={1\over 2}$ in Ref.~\cite{EKS}, where
the hamiltonian 
\begin{eqnarray}
    H={1\over 2
    }\int_{-\infty}^{0}\left[(\partial_{x}\phi)^{2}+
  \Pi^{2}\right]+v_{1}\cos\sqrt{2\pi\nu}\phi(0)\nonumber\\
 +v_{2}\cos\sqrt{2\pi/\nu}\phi(0)+v_{3}\Pi(0)^{2}\label{newham}
    \end{eqnarray}
was considered. Here, the $v_1$-term corresponds to tunneling of LQPs, the
    $v_2$-term to tunneling of electrons, and the $v_3$-term to an on-site
    density-density interaction on the constriction. The on-site
    density-density interaction reveals the reasonable assumption that
    electron forward scattering is stronger on the constriction than in the
    bulk. The basis of quasiparticles is the same as before, and it was shown
    in Ref.~\cite{EKS} that the only change is the boundary scattering matrix,
    which follows from 
\begeqar
P+Q&=&i\coth\left({\theta-\theta_{B}^{3}\over 2}-{i\pi\over
4}\right) \\
P-Q&=&-i\tanh\left({\theta-\theta_{B}^{1}\over 2}-{i\pi\over
4}\right)\coth\left({\theta-\theta_{B}^{2}\over 2}-{i\pi\over
4}\right) \nonumber
\endeqar
The relation between the bare parameters $v_{i}$ and the parameters
in the scattering matrix $\theta_{B}^{i}$ is a bit complicated (it also
depends on the regularization procedure adopted). However, it
simplifies in the limit of small bare couplings where we find, up
to irrelevant numerical factors, $v_{1}\propto e^{\theta_{B}^{1}/2}$,
$v_{2}\propto \left(e^{-\theta_{B}^{2}}+e^{-\theta_{B}^{3}}\right)$,
and $v_{3}\propto \left(e^{-\theta_{B}^{2}}-e^{-\theta_{B}^{3}}\right)$. In
the case $\nu=1/2$, the particles have a trivial interaction in the bulk, 
so the density is simply $\rho=e^{\theta}$, and the Fermi momentum
$e^{A}=V/2$. Assuming bare couplings are all rather small, we restrict to the
regime $\theta_B^1<<\theta_B^2,\theta_B^3$.

The interplay of the different terms can lead to several kinds of 
 behaviors. We only discuss here the question of effective tunneling
 charges. Thus, the curves representing the noise $S/V$ as a function of the
 current $I/V$ are considered. With a single tunneling term, see
 Eq.~(\ref{ham}), these curves have a single branch going from the origin to
 the point $I/V=\nu$. The most interesting feature of such a graph is its
 slope at the two extremities: near $I/V=0$ the slope is unity
 corresponding to tunneling of electrons, and near $I/V=\nu$ the slope is
 $(-\nu)$ corresponding to tunneling of LQPs with charge $\nu$. 

If we include irrelevant terms under the choice that $\theta_B^2 \approx
\theta_B^3$, the results are similar to the ones obtained with a single
tunneling term. This can be explicitely seen by considering a particle with
energy $e^{\theta}$; the boundary conditions it will see depend on the
magnitude of $\theta$, giving rise to two different energy domains:
\begin{equation}
    \begin{array}
{ccccccccc}
 &~&  \theta_{B}^{3}~~&\approx&~~\theta_{B}^{2}~~&\gg&~~\theta_{B}^{1}&~&\cr
    &P\approx 0&~&& ~&P\approx 0& ~&P\approx -i&\cr
    &Q\approx i&~&&~&Q\approx -i&~& Q\approx 0&\cr
    &N&~&&~&N&~&D&
    \end{array}
    \end{equation}
As the energy scale is increased - corresponding, for
instance, to the increase of the applied voltage - the system
successively sees D, then N  boundary conditions, just like in the case with a
single perturbation. Irrelevant terms do not have any strong qualitative
effect. 

It turns however out that things are quite different if there is a `resonance' between the $\cos\sqrt{2\pi/\nu}\phi$ and $(\Pi)^2$ terms. This corresponds to 
$\theta_{B}^{3} \gg \theta_{B}^{2} \gg \theta_{B}^{1}$ (similar results are
also obtained when $\theta_B^2\gg\theta_B^3\gg\theta_B^1$.) The meaning of the `resonance' between the two irrelevant terms is not entirely clear: technically, it corresponds to having the term of dimension 2 being the 
regularized square of the bare tunneling term, as $\partial\cos\sqrt{\pi}
\phi\cos\sqrt{\pi}\phi\propto \cos2\sqrt{\pi}\phi+(\partial\phi)^2$. In that
case, a variety of behaviors appears between the two energy scales
$e^{\theta_B^2}$ and $e^{\theta_B^3}$ and strongly counterintuitive
properties can be observed. If we consider again a particle with energy
$e^{\theta}$, the boundary conditions it will see now  fall within  four
energy domains:
\begin{equation}
    \begin{array}
{ccccccccc}
 &~&  \theta_{B}^{3}~~&\gg&~~\theta_{B}^{2}~~&\gg&~~\theta_{B}^{1}&~&\cr
    &P\approx 0&~&P\approx -i& ~&P\approx 0& ~&P\approx -i&\cr
    &Q\approx i&~& Q\approx 0&~&Q\approx -i&~& Q\approx 0&\cr
    &N&~&D&~&N&~&D&
    \end{array}
    \end{equation}
Consequently, as the energy scale is increased, the system
successively sees D, then N, then D, then N boundary conditions.

In the presence of the three tunneling terms, and  restricting
to the situation $\theta_{B}^{3} \gg \theta_{B}^{2} \gg \theta_{B}^{1}$ , the
current upon increasing the voltage increases from $0$ to a value very close 
to $I/V \approx \nu$, then goes back to a value very close  to $I/V \approx
0$, then back  to $I/V=\nu$. Only the two extremes are truly reached, and
correspond to the fixed points that were mentioned before. 

\begin{figure}
\vspace{0.3cm}
\begin{center}
\epsfig{file=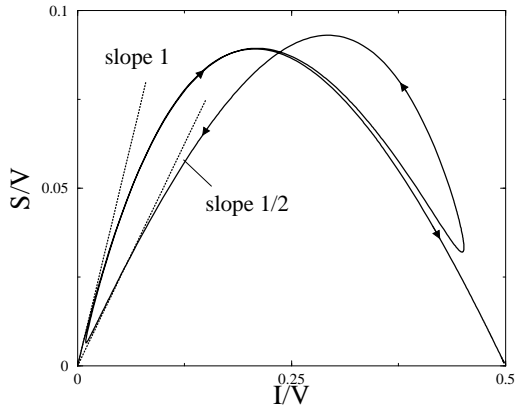,scale=0.4}  
\caption{\label{fig1}Shot noise vs. current for $\nu=1/2$ and effective
impurity strengths $(\t_B^1=\ln(1.01)) \ll (\t_B^2=\ln(100)) \ll
(\t_B^3=\ln(100000))$. The applied voltage $V$ is increased along the
direction of the arrows on the `roaming trajectory'. Evidently, the
effective charge exhibits changes as a function of the applied voltage.}
\end{center}
\end{figure}

Of special interest are then the evolution of the noise, and
especially the {\sl slopes} of $S$ versus $I$ near the fixed
points, as illustrated in Fig.~\ref{fig1}. 
Of course, a slope unity is observed at a real low energy $V \ll \theta_{B}^{1}$. In that case, all particles in the reservoir see 
D boundary conditions, $|Q|^{2}$ is always very small, $|Q|^4$
negligible compared to it. While the $N$ fixed point is initially approached
with a slope $(-\nu)$ (corresponding to the tunneling of LQPs) it is left,
as the voltage is increased, with a slope $(-1)$ instead of $(-\nu)$. This
conforms to the tunneling process of $1/\nu$ LQPs from one edge to the other
and was recently observed by Chung {\em et al.}~\cite{ComfortiI}. 

As the voltage is increased ever further and the strong backscattering fixed
point is approached again, the slope is equal to $\nu$ and not unity. In
a naive perturbative analysis, one would study the vicinity of the strong
backscattering fixed point and conclude that such a slope is not possible
since there is no corresponding operator - physically, LQPs cannot tunnel
`out' of the quantum Hall fluid, where only electrons are known to
exist. However, one should not forget that the strong backscattering fixed point is never actually 
reached on our trajectory - and presumably, what happens is that the
slope $\nu$ cannot be understood perturbatively in the vicinity of that
fixed point. Rather, in this case $\nu=1/2$, there are indications that 
the RG trajectory gets close to another fixed point, characterized by D
boundary conditions with $\phi$ pinned to values $\phi(0)={2n\pi\over
\sqrt{4\pi}}$. The slope $1/2$ could then be interpreted perturbatively in the
limit of {\sl that} fixed point. The situation bears some resemblance to the
case of resonant tunneling discussed in Ref.~\cite{KFold}, where the slope would be
interpreted as resulting from the tunneling of `half electrons', that is
(whole) electrons tunneling through an island. However, the situation is
different here: the hamiltonian is not the resonant tunneling hamiltonian. For
$\nu=1/2$, the resonant tunneling term $\cos 2\sqrt{2\pi\nu}\phi(0)$ is
irrelevant and would not lead to the phenomena described above. Nevertheless,
our findings might be interpreted in a way that the scattering potential effectively creates an `island' at intermediate energy scales.

\begin{figure}
\vspace{0.3cm}
\begin{center}
\epsfig{file=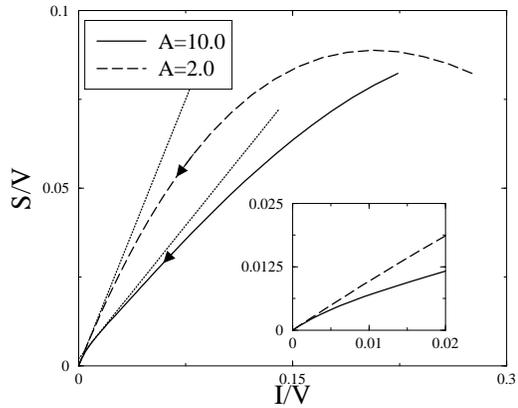,scale=0.4}  
\caption{\label{fig2}Shot noise vs. current for $\nu=1/2$ and fixed
voltages. The effective impurity strengths are varied as $\t_B^1=\t$,
$\t_B^2=100-\t$, and $\t_B^3=1000-10 \t$ with $\t \in [1.0,5.0]$. The inset
zooms into the region of very strong backscattering.}
\end{center}
\end{figure}

A drawback of the foregoing scenario is that, to observe the slope
$\nu$  at fixed $\theta_{B}$'s, the strong backscattering limit has to be
approached along an increasing voltage. This can easily be remedied
however. One can for instance consider instead fixing $V$, and
changing the $\theta_{B}$'s - corresponding to changing the
backscattering potential in an experiment. Fig.~\ref{fig2} shows the noise
and current for a particular trajectory where $\theta_{B}^{1}$ is increased
while $\theta_{B}^{2}$ and $\theta_{B}^{3}$ are decreased. 
This corresponds to increasing simultaneously the three
scattering terms  - hence, `pinching the point contact'. In that case, we find that, depending on the voltage, one can apparently 
 approach the strong backscattering limit with a slope
$\nu=1/2$. As emphasized in the inset of Fig.~\ref{fig2}, in the {\sl very
strong} backscattering limit ($S/V,I/V \approx 0.005$) the slope
$\nu=1/2$  reverts to a slope unity, the expected perturbative result. What we
see is that, for an appropriate choice of parameters, the domain where this
perturbative result can be observed is very small, while it is possible to get
instead a slope $\nu=1/2$ over a large domain, and very close to the strong backscattering limit.

Although the foregoing results are obtained for $\nu=1/2$, they should reflect
general properties of the {\sl field theory}
 for all values of  $\nu\leq {1\over 2}$, that is in the ``attractive regime'' of the boundary sine-Gordon model. We thus expect that these properties will be observed 
at all filling fractions $\nu={1\over t}$ with $t>1$ an
odd integer. Since these are the filling factors, where chiral edge
excitations are responsible for transport in real FQHE devices, the field
theories (\ref{ham}) and (\ref{newham}) provide a valid description of the
tunneling phenomena in such systems.

\begin{figure}
\vspace{0.3cm}
\begin{center}
\epsfig{file=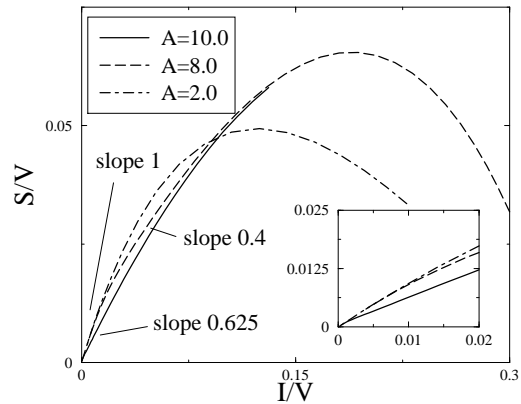,scale=0.4}  
\caption{\label{fig3}Shot noise vs. current for $\nu=1/3$ and fixed
voltages. The effective impurity strengths are varied as $\t_B^1=\t$,
$\t_B^2=100-\t$, and $\t_B^3=1000-10 \t$ with $\t \in [1.0,5.0]$. The inset
zooms into the region of very strong backscattering.}
\end{center}
\end{figure}

In any case, it is possible to carry out, to some extent, an analysis  similar to the 
 $\nu=1/2$ one for arbitrary $\nu$, based on the fact that the hamiltonian (\ref{newham}) is  integrable. 
A minor difference is that we do not know the correspondence between the
hamiltonian and the reflection matrix, which was afforded by the free 
fermion representation in the case $\nu=1/2$ . A more important difference is that,
for $\nu<1/2$, the second cosine is not the least
irrelevant one, which makes the integrable hamiltonian rather non generic. Now, for hamiltonian (\ref{newham}), general considerations ensure that the
reflection matrix differs from the usual ones by CDD phase factors (that is,
factors which preserve all the symmetries and physical properties of the
scattering matrices \cite{GhoZam}), and therefore must have the general form (recall $\gamma={1\over\nu}-1$):
\begeqar
P+Q&=&-\exp\left[i\chi_g(\theta)\right]\coth\left({\theta-\theta_{B}^{3}\over 2}-{i\pi\over
4}\right) \nonumber\\
P-Q&=&
\exp\left[i\chi_g(\theta)\right]\tanh\left({\gamma(\theta-\theta_{B}^{1})\over
2} -{i\pi\over 4}\right)\nonumber\\
&\times&\coth\left({\theta-\theta_{B}^{2}\over 2}-{i\pi\over
4}\right) .
\endeqar
The transmission probability $|Q|^{2}$ can then be extracted from these
expressions, and current and noise computed as for $\nu=1/2$. Similar results
are observed, with the main difference that the slopes near the weak and
strong backscattering limit are not restricted to unity and $\nu$ any
longer: intermediate values can appear, {\em cf.} Fig.~\ref{fig3}.  
Remarkably, slopes varying between $1/3$ and unity were observed in the
noise experiments on FQHE devices at filling fraction $\nu=1/3$ in
Ref.~\cite{Comforti}.     

In conclusion, we have shown that the inclusion of irrelevant terms of a
reasonable magnitude can lead to situations, where the effective charge
extracted from the slope of the shot noise takes counterintuitive
values. Although we have not precisely investigated the experimental scenario
of diluted beams incident on a quantum point contact, our predictions could
well be important to describe the so far unexplained experimental results in
Ref.~\cite{Comforti}. We believe that at intermediate energies irrelevant
operators are generally important. Therefore, non perturbative
analyses are essential to interpret transport experiments at finite
temperature and applied voltage.   

We thank  Y.~C.~Chung, R.~Egger, H.~Grabert,
M.~Heiblum, and C.~L.~Kane for helpful discussions. HS benefitted from the
kind hospitality of the   Freiburg Physics Department  while this work was
carried out, and was supported by the DOE and the Humboldt Foundation.  BT was
supported  by the DFG and the LSF program.


\end{document}